\documentclass[10pt, conference, compsocconf]{IEEEtran}
\usepackage{flushend}
\usepackage{ifpdf}
\usepackage{cite}
\ifCLASSINFOpdf
\else
\fi
\usepackage[cmex10]{amsmath}
\usepackage{algorithmic}
\usepackage{array}
\usepackage[tight,font=footnotesize]{subfig}
\usepackage{fixltx2e}
\usepackage{url}
\usepackage{xspace}
\usepackage{amssymb}
\usepackage{amsmath}
\usepackage{color}

\usepackage{pgf}
\usepackage{tikz}
\usepackage{forloop}
\usepackage{ifthen}
\usetikzlibrary{plotmarks}
\usetikzlibrary{patterns}
\usetikzlibrary{trees}
\usepackage{dblfloatfix}
\usepackage{todonotes}

\usepackage{multirow}
\usepackage{placeins}

\usepackage{macro}
\usepackage{float} %% To put figures at the correct place
\usepackage{tikz}
\usepackage{forloop}
\usepackage{ifthen}
\usepackage{subfig}
\usepackage{lscape}
\usepackage{pgfpages}
\usepackage{pgfplotstable}
\usepackage{pgfplots}
\usepackage{rotating}
\usepackage{tikz}
\usepackage{xstring}
\usepackage{booktabs}
\usepackage{array}
\usepackage{longtable}
\usepackage{xargs}
\pgfplotsset{compat=1.5.1}

\usetikzlibrary{spy, calc}

\pgfplotsset{
  discard if not/.style 2 args={
    x filter/.code={
      \edef\tempa{\thisrow{#1}}
      \edef\tempb{#2}
      \ifx\tempa\tempb
      \else
      \def\pgfmathresult{inf}
      \fi
    }
  }
}

\makeatletter
\pgfplotstableset{
  discard if not/.style 2 args={
    row predicate/.code={
      \def\pgfplotstable@loc@TMPd{\pgfplotstablegetelem{##1}{#1}\of}
      \expandafter\pgfplotstable@loc@TMPd\pgfplotstablename
      \edef\tempa{\pgfplotsretval}
      \edef\tempb{#2}
      \ifx\tempa\tempb
      \else
      \pgfplotstableuserowfalse
      \fi
    }
  }
}
\makeatother

\makeatletter
\pgfplotstableset{
    discard if notmatsolv/.style 2 args={
        row predicate/.code={
            \def\pgfplotstable@loc@TMPd{\pgfplotstablegetelem{##1}{MATRIX}\of}
            \expandafter\pgfplotstable@loc@TMPd\pgfplotstablename
            \edef\tempa{\pgfplotsretval}
            \def\pgfplotstable@loc@TMPe{\pgfplotstablegetelem{##1}{SOLVER}\of}
            \expandafter\pgfplotstable@loc@TMPe\pgfplotstablename
            \edef\tempb{\pgfplotsretval}
            \edef\tempc{#1}
            \edef\tempd{#2}
            \ifx\tempa\tempc
            \ifx\tempb\tempd
            \else
                \pgfplotstableuserowfalse
            \fi
            \else
                \pgfplotstableuserowfalse
            \fi

        }
    }
}
\makeatother

\pgfplotsset{
    discard if notmatsolv/.style 2 args={
        x filter/.code={
            \edef\tempa{\thisrow{MATRIX}}
            \edef\tempb{#1}
            \ifx\tempa\tempb

            \edef\tempc{\thisrow{SOLVER}}
            \edef\tempd{#2}
            \ifx\tempc\tempd

            \edef\tempc{\thisrow{NCUDA}}
            \edef\tempd{0}
            \ifx\tempc\tempd
            \else
            \def\pgfmathresult{inf}
            \fi

            \else
            \def\pgfmathresult{inf}
            \fi
            
            \else
            \def\pgfmathresult{inf}
            \fi
        }
    }
}
\makeatother

\makeatletter

\pgfplotsset{
    discard if notmatsolvfull/.style 2 args={
        x filter/.code={
            \edef\tempa{\thisrow{MATRIX}}
            \edef\tempb{#1}
            \ifx\tempa\tempb

            \edef\tempc{\thisrow{SOLVER}}
            \edef\tempd{#2}
            \ifx\tempc\tempd

            \edef\tempc{\thisrow{NCUDA}}
            \edef\tempd{\thisrow{NCPU}}
            \edef\tempe{0}
            \edef\tempf{12}
            \ifx\tempd\tempf
            \else
            \ifx\tempc\tempe
            \def\pgfmathresult{inf}
            \else
            \fi
            \fi

            \else
            \def\pgfmathresult{inf}
            \fi
            
            \else
            \def\pgfmathresult{inf}
            \fi
        }
    }
}

\newcommand{\pgfplotstablefilterrows}[3]
{
  \pgfplotstablegetrowsof{#1}
  \pgfmathsetmacro{\NumOfRows}{\pgfplotsretval}
  \pgfmathsetmacro{\MaxRow}{\NumOfRows-1}
  \pgfplotstablegetcolsof{#1}
  \pgfmathsetmacro{\NumOfCols}{\pgfplotsretval}

  \pgfplotstabletranspose{\TransposedData}{#1}
  \pgfplotstableset{create on use/TransposedHead/.style={copy column from table={\TransposedData}{[index]0}}}
  \pgfplotstablenew[columns={TransposedHead}]{\NumOfCols}{\TransposedFilteredData}
  \pgfplotsforeachungrouped \pgfplotstablerowindex in {0,1,...,\MaxRow}{ % Row loop
    #3
  }
  \pgfplotstabletranspose[colnames from=TransposedHead,input colnames to=]{#2}{\TransposedFilteredData}
  \pgfplotstableclear{\TransposedData}
  \pgfplotstableclear{\TransposedFilteredData}
}

\newcommand{\pgfplotstablefilterrowsmatrix}[3]{
  \pgfplotstablefilterrows{#1}{#2}{
    \pgfplotstablegetelem{\pgfplotstablerowindex}{MATRIX}\of#1
    \IfStrEq{\pgfplotsretval}{#3}{
      \pgfplotstablecreatecol[copy column from table={\TransposedData}{\pgfplotstablerowindex}]{\pgfplotstablerowindex}{\TransposedFilteredData}
    }{}
  }
}

\newcommand{\pgfplotstablefilterrowsmatrixsolvercpu}[4]
{
  \pgfplotstablefilterrows{#1}{#2}{
    \pgfplotstablegetelem{\pgfplotstablerowindex}{MATRIX}\of#1
    \IfStrEq{\pgfplotsretval}{#3}{
      \pgfplotstablegetelem{\pgfplotstablerowindex}{SOLVER}\of#1
      \IfStrEq{\pgfplotsretval}{#4}{%TRUE
        \pgfplotstablegetelem{\pgfplotstablerowindex}{NCUDA}\of#1
        \IfStrEq{\pgfplotsretval}{0}{%TRUE
          \pgfplotstablecreatecol[copy column from table={\TransposedData}{\pgfplotstablerowindex}]{\pgfplotstablerowindex}{\TransposedFilteredData}
        }{}
      }{}
    }{}
  }
}

\newcommandx{\drawresultscpu}[4][1=axis,2=,4=results.data]{
  \begin{figure}
    \begin{tikzpicture}
      \begin{#1}[ylabel={Time [s]}, xlabel={Number of cores}, xtick={1,3,6,9,12}, #2]
        \addplot+[discard if notmatsolv={#3}{pastix}] table[x=NCPU, y=Facto] {#4};\addlegendentry{pastix}
        \addplot+[discard if notmatsolv={#3}{StarPU}] table[x=NCPU, y=Facto] {#4};\addlegendentry{StarPU}
        \addplot+[discard if notmatsolv={#3}{PARSEC}] table[x=NCPU, y=Facto] {#4};\addlegendentry{ParSEC}
      \end{#1}
    \end{tikzpicture}
    \caption{Factorization time for matrix #3}
  \end{figure}
}

\newcommandx{\drawresultscpuflops}[4][1=axis,2=,4=results.data]{
  \begin{figure}
    \begin{tikzpicture}
      \begin{#1}[ylabel={FLOPS}, xlabel={Number of cores}, xtick={1,3,6,9,12}, #2]
        \addplot+[discard if notmatsolv={#3}{pastix}] table[x=NCPU, y=FLOPS] {#4};\addlegendentry{pastix}
        \addplot+[discard if notmatsolv={#3}{StarPU}] table[x=NCPU, y=FLOPS] {#4};\addlegendentry{StarPU}
        \addplot+[discard if notmatsolv={#3}{PARSEC}] table[x=NCPU, y=FLOPS] {#4};\addlegendentry{ParSEC}
      \end{#1}
    \end{tikzpicture}
    \caption{Factorization FLOPS for matrix #3}
  \end{figure}
}

\newcommandx{\drawresultsgpu}[4][1=axis,2=,4=results.data]{
  \begin{figure}
    \begin{tikzpicture}
      \pgfplotsset{every axis legend/.append style={
          at={(0.5,1.03)},
          anchor=south}}
      \begin{#1}[ylabel={Time [s]}, xlabel={Number of GPU}, xtick={0,1,2,3}, #2]
        \addplot+[discard if notmatsolvfull={#3}{StarPU}] table[x=NCUDA, y=Facto] {#4};\addlegendentry{StarPU, 12 workers including GPUs}
        \addplot+[discard if notmatsolvfull={#3}{PARSEC}] table[x=NCUDA, y=Facto] {#4};\addlegendentry{ParSEC, 12 cores plus GPUs}
      \end{#1}
    \end{tikzpicture}
    \caption{Factorization time for matrix #3}
  \end{figure}
}

\newcommandx{\drawresultsgpuflops}[4][1=axis,2=,4=results.data]{
  \begin{figure}
    \begin{tikzpicture}
      \pgfplotsset{every axis legend/.append style={
          at={(0.5,1.03)},
          anchor=south}}
      \begin{#1}[ylabel={FLOPS}, xlabel={Number of GPU}, xtick={0,1,2,3}, #2]
        \addplot+[discard if notmatsolvfull={#3}{StarPU}] table[x=NCUDA, y=FLOPS] {#4};\addlegendentry{StarPU, 12 workers including GPUs}
        \addplot+[discard if notmatsolvfull={#3}{PARSEC}] table[x=NCUDA, y=FLOPS] {#4};\addlegendentry{ParSEC, 12 cores plus GPUs}
      \end{#1}
    \end{tikzpicture}
    \caption{Factorization FLOPS for matrix #3}
  \end{figure}
}

\newcommand{\pgfplotstablefilterrowsmatrixsolverfull}[4]
{
  \pgfplotstablefilterrows{#1}{#2}{
    \pgfplotstablegetelem{\pgfplotstablerowindex}{MATRIX}\of#1
    \IfStrEq{\pgfplotsretval}{#3}{
      \pgfplotstablegetelem{\pgfplotstablerowindex}{SOLVER}\of#1
      \IfStrEq{\pgfplotsretval}{#4}{%TRUE
        \pgfplotstablegetelem{\pgfplotstablerowindex}{NCUDA}\of#1
        \pgfmathtruncatemacro{\ncuda}{\pgfplotsretval}
        \pgfplotstablegetelem{\pgfplotstablerowindex}{NCPU}\of#1
        \pgfmathtruncatemacro{\ncpu}{\pgfplotsretval}
        \pgfmathtruncatemacro{\tempva}{\ncpu + \ncuda}
        \ifnum\tempva=12%true
        \pgfplotstablecreatecol[copy column from table={\TransposedData}{\pgfplotstablerowindex}]{\pgfplotstablerowindex}{\TransposedFilteredData}
        \else
        \fi
      }{}
    }{}
  }
}

\usetikzlibrary{plotmarks}
\usepgfplotslibrary{units}

\pgfplotsset{
  barchartstyle/.style={
    legend style={
      at={(0.5,0.95)}, anchor=south,
      column sep=10pt,
      font=\small,
    },
    ybar=1pt,
    ylabel=Performance (\gflops{}),
    bar width=2pt,
    axis on top,
    xtick=data,
    xticklabels={
      {afshell10(D, \LU{})},
      {FilterV2(Z, \LU{})},
      {Flan(D, \LLT{})},
      {audi(D, \LLT{})},
      {MHD(D, \LU{})},
      {Geo1438(D, \LLT{})},
      {pmlDF(Z, \LDLT{})},
      {HOOK(D, \LU{})},
      {Serena(D, \LDLT{})}
    },x tick label style={rotate=20,anchor=east},
  }
}

\begin{document}
%\title{Sparse direct solvers with accelerators over DAG runtimes}
\title{Taking advantage of hybrid systems for sparse direct solvers via task-based runtimes}
% \title{Using task-based runtimes to improve the performance of sparse
%   direct solvers over heterogeneous systems}

\author{  \IEEEauthorblockN{Xavier Lacoste, Mathieu Faverge,\\ Pierre Ramet, Samuel Thibault}
  \IEEEauthorblockA{INRIA - IPB - LaBRI\\ University of Bordeaux\\
    Talence, France\\
    Email: \{xavier.lacoste,mathieu.faverge\}@inria.fr,\\
           \{pierre.ramet,samuel.thibault\}@inria.fr}
  \and
  \IEEEauthorblockN{George Bosilca}
  \IEEEauthorblockA{Innovative Computing Laboratory\\
    University of Tennessee\\
    Knoxville, Tennessee, USA\\
    Email: \{bosilca\}@icl.utk.edu}
}


%Telephone: (800) 555--1212, Fax: (888) 555--1212}


\date{\today} %%{27 April 2012}

\maketitle

\begin{abstract}

The ongoing hardware evolution exhibits an escalation in the number,
as well as in the heterogeneity, of computing resources.  The pressure
to maintain reasonable levels of performance and portability forces
application developers to leave the traditional programming paradigms
and explore alternative solutions.  \pastix is a parallel sparse
direct solver, based on a dynamic scheduler for modern hierarchical
manycore architectures. In this paper, we study the benefits and
limits of replacing the highly specialized internal scheduler of the
\pastix solver with two generic runtime systems: \parsec and \starpu.
The tasks graph of the factorization step is made available to the two
runtimes, providing them the opportunity to process and optimize its
traversal in order to maximize the algorithm efficiency for the
targeted hardware platform.  A comparative study of the performance of
the \pastix solver on top of its native internal scheduler, \parsec{},
and \starpu{} frameworks, on different execution environments, is
performed.  The analysis highlights that these generic task-based
runtimes achieve comparable results to the application-optimized
embedded scheduler on homogeneous platforms. Furthermore, they are
able to significantly speed up the solver on heterogeneous
environments by taking advantage of the accelerators while hiding the
complexity of their efficient manipulation from the programmer.

\end{abstract}

%% \begin{IEEEkeywords}

%% \end{IEEEkeywords}

% \fi
\IEEEpeerreviewmaketitle

\section{Introduction}
\label{sec:intro}

Emerging processor technologies put an emphasis on increasing the
number of computing units instead of increasing their working
frequencies. As a direct outcome of the physical multiplexing of
hardware resources, complex memory hierarchies had to be instated to
relax the memory bottleneck and ensure a decent rate of memory
bandwidth for each resource. The memory becomes divided in several
independent areas, capable of delivering data simultaneously through a
complex and hierarchical topology, leading to the mainstream Non
Uniform Memory Accesses (NUMA) machines we know today.  Together with
the availability of hardware accelerators, this trend profoundly
altered the execution model of current and future platforms,
progressing them toward a scale and a complexity unattained before.
Furthermore, with the established integration of accelerators into
modern architectures, such as \gpu{}s or Intel Xeon Phis, high-end
multi-core \cpu{}s are consistently outperformed by these novel, more
integrated, architectures both in terms of data processing rate and
memory bandwidth. As a consequence, the working environment of today's
application developers evolved toward a multi-level massively parallel
environment, where computation becomes cheap but data movements
expensive, driving sup the energetic cost and algorithmic overheads
and complexities.

With the advent of APIs for \gpu{} programming, such as \cuda{} or
OpenCL, programming accelerators has been rapidly evolving in the past
years, permanently bringing accelerators into the mainstream.  Hence,
\gpu{}s are becoming a more and more attractive alternative to
traditional \cpu{}s, particularly for their more interesting
cost-per-flop and watts-per-flop ratios. However, the availability of
a particular programming API only partially addresses the development
of hybrid algorithms capable of taking advantage of all computational
resources available, including accelerators and \cpu{}s.  Extracting a
satisfactory level of performance, out of such entangled
architectures, remains a real challenge due to the lack of consistent
programming models and tools to assess their performance. In order to
efficiently exploit current and future architectures, algorithm
developers are required to expose a large amount of parallelism, adapt
their code to new architectures with different programming models, and
finally, map it efficiently on the heterogeneous resources. This is a
gargantuan task for most developers as they do not possess the
architectural knowledge necessary to mold their algorithms on the
hardware capabilities in order to achieve good efficiency, and/or do
not want to spend new efforts with every new generation of hardware.

Solving large sparse linear systems of equations, $Ax=b$, is one of
the most important and time-consuming parts in many scientific and
engineering algorithms, a building block toward more complex
scientific applications.  A significant amount of research has been
done on dense linear algebra, but sparse linear algebra on
heterogeneous system is a work-in-progress.
Multiple reasons warrant this divergence, including the intrinsic
algorithmic complexity and the highly irregular nature of the
resulting problem, both in terms of memory accesses and computational
intensities. Combined with the heterogeneous features of current and
future parallel architectures, this depicts an extremely complex
software development field.

The \pastix solver is a sparse direct solver that can solve symmetric
definite, indefinite, and general problems using Cholesky, \LDLT{},
and \LU{} factorizations, respectively.  The \pastix implementation
relies on a two-level approach using the POSIX Thread library within a
node, and the Message Passing Interface (MPI) between different nodes.
Historically, \pastix scheduling strategy was based on a cost model of
the tasks executed that defines the execution order used at runtime
during the analyze phase. In order to complement the lack of precision
of the cost model on hierarchical architectures, a dynamic scheduler
based on a work-stealing strategy has been developed to reduce the
idle times while preserving a good locality for data
mapping~\cite{C:LaBRI::siam2012}.  More recently, the solver has been
optimized to deal with the new hierarchical multi-core
architectures~\cite{faverge:inria-00344026}, at the level of internal
data structures of the solver, communication patterns, and scheduling
strategies.

In this paper, we advance the state-of-the-art in supernodal solvers
by migrating \pastix toward a new programming paradigm, one with a
promise of efficiently handling hybrid execution environments while
abstracting the application from the hardware constraints. Many
challenges had to be overcome, going from exposing the \pastix
algorithms using a task-based programming paradigm, to delivering a
level of task parallelism, granularity, and implementation allowing
the runtime to efficiently schedule the resulting, highly irregular
tasks, in a way that minimizes the execution span.
We exposed the original algorithm using the concept of tasks, a
self-contained computational entity, linked to the other tasks by data
dependencies. Specialized task-graph description formats were used in
accordance with the underlying runtime system (\parsec or \starpu). We
provided specialized \gpu{}-aware versions for some of the most
compute intensive tasks, providing the runtimes with the opportunity
to unroll the graph of tasks on all available computing resources. The
resulting software is, to the best of our knowledge, the first
implementation of a sparse direct solver with a supernodal method
supporting hybrid execution environments composed of multi-cores and
multi-GPUs. Based on these elements, we pursue the evaluation of the
usability and the appeal of using a task-based runtime as a substrate
for executing this particular type of algorithm, an extremely
computationally challenging sparse direct solver. Furthermore, we take
advantage of the integration of accelerators (\gpu{}s in this context)
with our supporting runtimes, to evaluate and understand the impact of
this drastically novel portable way of writing efficient and perennial
algorithms.
Since the runtime system offers a uniform programming interface,
dissociated from a specific set of hardware or low-level software
entities, applications can take advantage of these uniform programming
interfaces for ensuring their portability.  Moreover, the exposed
graph of tasks allows the runtime system to apply specialized
optimization techniques and minimize the application's time to
solution by strategically mapping the tasks onto computing resources
by using state-of-the-art scheduling strategies.

The rest of the paper is organized as follows. We describe the
supernodal method of the \pastix solver in Section~\ref{sec:pastix},
followed by a description of the runtimes used~\ref{sec:scheduler}.
Section~\ref{sec:implementation} explains the implementation over the
DAG schedulers with a preliminary study over multi-core architectures,
followed by details on the extension to heterogeneous
architectures. All choices are supported and validated by a set of
experiments on a set of matrices with a wide range of characteristics.
Finally, section~\ref{sec:conclusion} concludes with some prospects of
the current work.

\section{Related Work}
\label{sec:related}

The dense linear algebra community spent a great deal of effort to tackle
the challenges raised by the sharp increase of the number of computational
resources. Due to their heavy computational cost, most of their algorithms
are relatively simple to handle. Avoiding common pitfalls such as the
``fork-join'' parallelism, and expertly selecting the blocking factor, provide an almost
straightforward way to increase the parallelism and thus achieve better
performance. Moreover, due to their natural load-balance, most of the algorithms
can be approached hierarchically, first at the node level, and then at the
computational resource level. In a shared memory context, one of the seminal
papers~\cite{icl:369} replaced the commonly used LAPACK layout with one
based on tiles/blocks. Using basic operations on these tiles exposes the
algorithms as a graph of tasks augmented with dependencies between them. In
shared memory, this approach quickly generates a large number of ready tasks,
while, in distributed memory, the dependencies allow the removal of hard
synchronizations. This idea leads to the design of new algorithms for various
algebraic operations \cite{icl:509}, now at the base of well-known
software packages like PLASMA~\cite{icl:486}.

This idea is recurrent in almost all novel approaches surrounding the many-core
revolution, spreading outside the boundaries of dense linear algebra. Looking at
the sparse linear algebra, the efforts were directed toward improving the
behavior of the existing solvers by taking into account both task and data
affinity and relying on a two-level hybrid parallelization approach, mixing
multithreading and message passing. Numerous solvers are now able to efficiently
exploit the capabilities of these new
platforms~\cite{faverge:inria-00344026,pardiso}.  New solvers have also
been designed and implemented to address these new platforms. For them the
chosen approach follows the one for dense linear algebra, fine-grained
parallelism, thread-based parallelization, and advanced data management to deal
with complex memory hierarchies. Examples of this kind of solver are
HSL~\cite{hogg2010design} and SuperLU-MT~\cite{li:08} for sparse LU or
Cholesky factorizations and SPQR~\cite{davis:11} and
\texttt{qr\_mumps}~\cite{butt:11} for sparse QR factorizations.

With the advent of accelerator-based platforms, a lot of attention has shifted
toward extending the multi-core aware algorithms to fully exploit the huge
potential of accelerators (mostly \gpu{}s). The main challenges raised by these
heterogeneous platforms are mostly related to task granularity and data
management: although regular cores require fine granularity of data as well as
computations, accelerators such as GPUs need coarse-grain tasks. This inevitably
introduces the need for identifying the parts of the algorithm which are more
suitable to be processed by accelerators. As for the multi-core case described
above, the exploitation of this kind of platform was first considered in the
context of dense linear algebra algorithms.

Moreover, one constant becomes clear: a need for a portable layer that will
insulate the algorithms and their developers from the rapid hardware
changes. Recently, this portability layer appeared under the denomination of
a task-based runtime. The algorithms are described as tasks with data dependencies
in-between, and the runtime systems are used to manage the tasks dynamically and
schedule them on all available resources. These runtimes can be generic, like
the two runtimes used in the context of this study
(\starpu~\cite{Augonnet_2010_ccpe} or \parsec~\cite{dague:12}), or more specialized
like QUARK~\cite{icl:759}.
These efforts resulted in the design of the DPLASMA
library~\cite{dague-la} on top of \parsec and the adaptation of the existing FLAME
library~\cite{flamegpu}.
On the sparse direct methods front, preliminary work has resulted in mono-GPU
implementations based on offloading parts of the computations to the
GPU~\cite{George2011,Lucas:2010,Yu2011}. Due to its very good data locality,
the multifrontal method is the main target of these approaches. The main idea is
to treat some parts of the task dependency graph entirely on the GPU. Therefore,
the main originality of these efforts is in the methods and algorithms used to
decide whether or not a task can be processed on a GPU. In most cases, this was
achieved through a threshold based criterion on the size of the computational
tasks.

Many initiatives have emerged in previous years to develop efficient runtime
systems for modern heterogeneous platforms. Most of these runtime systems use a
task-based paradigm to express concurrency and dependencies by employing a task
dependency graph to represent the application to be executed. Without going into
details, the main differences between these approaches are related to their
representation of the graph of tasks, whether they manage data movements between
computational resources, the extent they focus on task scheduling, and their
capabilities to handle distributed platforms.

\section{Supernodal factorization}
\label{sec:pastix}

Sparse direct solvers are algorithms that address sparse matrices,
mostly filled with zeroes.  In order to reduce the
number of operations, they consider only non-zeroes of the matrix $A$.
During factorization, new non-zero entries -- called fill-in -- appear
in the factorized matrix and lead to more computation and memory
consumption. One of the main objectives of those
solvers is to keep the fill-in to its minimum to limit the memory
consumption. 
The first step of any sparse direct solver is the computation of a
nested dissection of the problem that results in a permutation of the
unknowns of $A$. This process is a graph partitioning algorithm
applied to the connectivity graph associated with the matrix.
The computed permutation reduces the fill-in that the factorization
process will generate, and the elimination tree~\cite{r87} is
generated out of the
separators discovered during the graph partitioning process.
Basically, each node of the tree represents the set of unknowns that
belongs to a separator, and edges are connections between those
separators in the original graph. The edges of the tree connect a node
to almost all nodes in the path that connect it to the root of the
tree. They represent contributions from one node to another during the
factorization.
The second step of sparse solvers is the analysis stage which predicts
the non-zero pattern of the factorized matrix through a symbolic
factorization. The resulting pattern is grouped in blocks of non-zero
elements to allow for more efficient \blas calls. Blocks can be
enlarged, if extra fill-in is allowed, for better performance, or
split to generate more parallelism.

\begin{figure}[!t]
  \centering
  \includegraphics[width=0.35\textwidth]{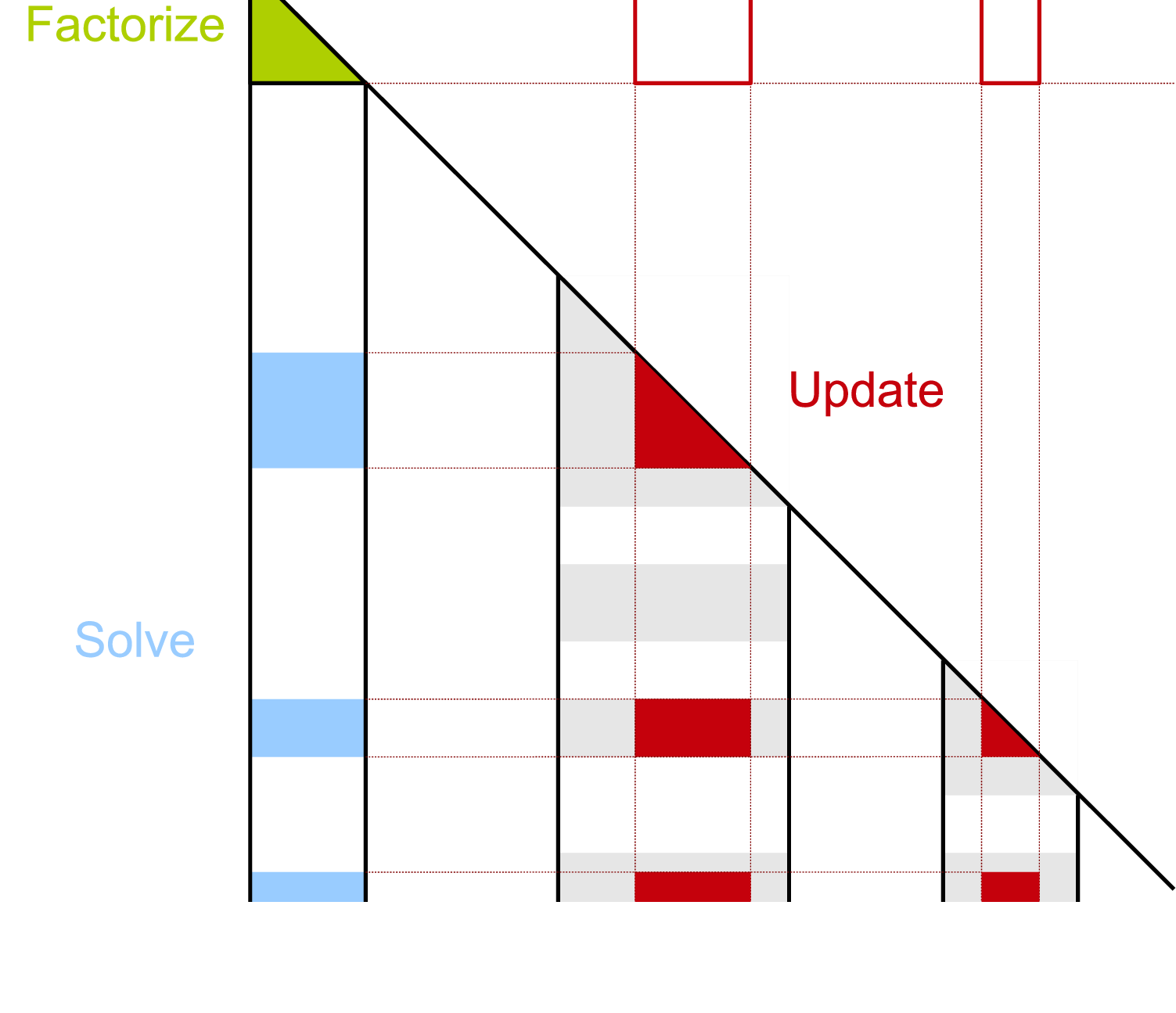} 
  \caption{Decomposition of the task applied while processing one panel}
  \label{fig:update}
\end{figure}

Once the elimination tree and the symbolic factorization are computed,
two different methods can be applied:
multifrontal~\cite{duff1983multifrontal} or
supernodal~\cite{ashcraft1987progress}.
The \pastix{} solver uses a supernodal method. 
Each node of the elimination tree, or supernode, represents a subset
of contiguous columns, also called a panel, in the matrix.
To each node of the elimination tree, we associate a task, called 1D
task, that performs three steps
associated with the panel $A$, as shown in the Figure~\ref{fig:update}:
\begin{enumerate}
\item Factorization of the diagonal block,
\item Triangular solve on the off-diagonal blocks in the panel (\trsm), and
\item For each off-diagonal block $A_i$, apply the associated update to the
  facing panel $C_i$ (\gemm{}) -- we call $C_i$ the facing panel with
  the diagonal block owning the same rows as the off-diagonal block $A_{i}$.
\end{enumerate}

Figure~\ref{fig:update} represents a lower triangular matrix used in
case of symmetric problems with Cholesky factorization, but is also
valid for non-symmetric cases with \pastix. In a general manner,
\pastix works on the matrix $A+A^T$, which produces a symmetric
pattern. In non-symmetric cases, the steps 2 and 3 are then 
duplicated for the $L$ and $U$ matrices of the \LU{}
factorization. Besides, the \pastix solver doesn't perform dynamic
pivoting, as opposed to SuperLU~\cite{superlu99}, which allows the
factorized matrix structure to be fully known at the analysis step.
In the rest of the paper, we will discuss only the Cholesky
implementation. \LDLT{} and \LU{} factorizations follow the same method.

The \pastix solver relies on a cost model of this 1D task to compute a static
scheduling.
This static scheduling associates ready tasks with the first available
resources among the computational units. The complexity of such an
algorithm depends on the number of tasks and resources. Hence, the
1D task is kept as a large single task to lower the complexity
of the analysis part.
However, it is obvious that more parallelism could be extracted from
those tasks, but would increase the analysis step complexity.

First, the triangular solves, applied on off-diagonal blocks of each
panel, are independent computations that depend only on the
diagonal block factorization. Thus, each panel is stored as a single
tall and skinny matrix, such that the \trsm granularity can be
decided at runtime and is independent of the data storage.
At lower levels of the elimination tree, the small block granularity
might induce a large overhead if they are considered as independent
tasks. On the contrary, at higher levels, the larger supernodes (Order
of $N^{\frac{2}{3}}$ for a 3D problem of $N$ unknowns, or $sqrt(N)$
for a 2D problem) might be split to create more parallelism with low
overhead. That is why supernodes of the higher levels are split
vertically prior to the factorization to limit the task granularity
and create more parallelism. In this study, to compare to the
existing \pastix solver, we keep all \trsm{} operations on a single
panel grouped together as a single operation.

Second, the same remark as before applies to the update tasks with a
higher order of magnitude as before. Each couple of off-diagonal blocks $(A_i, A_j)$, with $i<j$ in a panel,
generates an independent update to the trailing submatrix formed by their outer
product. To adapt to the small granularity of off-diagonal blocks in sparse
solvers, those updates are grouped together. Two variants
exists. \emph{Left-looking}: all tasks contributing to a single panel are
associated in a single task, they have a lot of input edges and only one in-out
data. \emph{Right-looking}: all updates generated by a single panel are directly
applied to the multiple destination panels. This solution has a single input
data, and many panels are accessed as in-out. \pastix uses the
\emph{right-looking} variant.  Therefore, the nature of the supernodal algorithm
is in itself a DAG of tasks dependent on the structure of the factorized matrix, but
independent of the numerical content thanks to the static pivoting strategy.
However, since we consider a data as a panel,
and both of the targeted runtime systems take a fixed number of dependencies per
tasks, one update task will be generated per couple of panels instead of one per
panel, as in \pastix.

\section{Runtimes}
\label{sec:scheduler}

In our exploratory approach toward moving to a generic scheduler for \pastix, we
considered two different runtimes: \starpu and \parsec. Both runtimes have been proven
mature enough in the context of dense linear algebra, while providing two
orthogonal approaches to task-based systems.

The \parsec~\cite{dague:12} distributed runtime system is a generic data-flow
engine supporting a task-based implementation targeting hybrid
systems. Domain specific languages are available to expose a user-friendly
interface to developers and allow them to describe their algorithm using
high-level concepts. This programming paradigm constructs an abridged
representation of the tasks and their dependencies as a graph of tasks -- a
structure agnostic to algorithmic subtleties, where all intrinsic knowledge
about the complexity of the underlying algorithm is extricated, and the only
constraints remaining are annotated dependencies between tasks~\cite{CJY04}.
This symbolic representation, augmented with a specific data distribution, is then
mapped on a particular execution environment. The runtime supports the usage of
different types of accelerators, \gpu{}s and Intel Xeon Phi, in addition to
distributed multi-core processors. Data are transferred between computational
resources based on coherence protocols and computational needs, with emphasis on
minimizing the unnecessary transfers. The resulting tasks are dynamically scheduled on
the available resources following a data reuse policy mixed with
different criteria for adaptive scheduling. The entire runtime targets very fine
grain tasks (order of magnitude under ten microseconds), with a flexible
scheduling and adaptive policies to mitigate the effect of system noise and take
advantage of the algorithmic-inherent parallelism to minimize the execution
span.

The experiment presented in this paper takes advantage of a specialized domain
specific language of \parsec, designed for affine loops-based
programming~\cite{dague-la}. This specialized interface allows for a drastic
reduction in the memory used by the runtime, as tasks do not exist until they
are ready to be executed, and the concise representation of the task-graph allows
for an easy and stateless exploration of the graph. In exchange for the memory
saving, generating a task requires some extra computation, and lies in the
critical path of the algorithm. The need for a window of visible tasks is then
pointless, the runtime can explore the graph dynamically based on the ongoing
state of the execution.

%\subsection{\starpu{}}

\starpu~\cite{Augonnet_2010_ccpe} is a runtime system aiming to allow
programmers to exploit the computing power of clusters of hybrid systems
composed of \cpu{}s and various accelerators (\gpu{}s, Intel Xeon Phi, etc)
while relieving them from the need to specially adapt their programs to the
target machine and processing units.  The \starpu{} runtime supports a {\em
  task-based programming model}, where applications submit computational tasks,
with \cpu{} and/or accelerator implementations, and \starpu{} schedules these
tasks and associated data transfers on available \cpu{}s and accelerators. The
data that a task manipulates is automatically transferred among accelerators and
the main memory in an optimized way (minimized data transfers, data prefetch,
communication overlapped with computations, etc.), so that programmers are
relieved of scheduling issues and technical details associated with these
transfers. \starpu{} takes particular care of scheduling tasks efficiently, by
establishing performance models of the tasks through on-line measurements, and
then using well-known scheduling algorithms from the literature.  In addition, it allows
scheduling experts, such as compilers or computational library developers, to
implement custom scheduling policies in a portable fashion.

The differences between the two runtimes can be classified into two groups:
conceptual and practical differences.  At the conceptual level 
the main differences between \parsec and \starpu are the tasks submission
process, the centralized scheduling, and the data movement strategy. \parsec uses
its own parameterized language to describe the DAG in comparison with the simple
sequential submission loops typically used with \starpu. Therefore, \starpu relies on a
centralized strategy that analyzes, at runtime, the dependencies between tasks and schedules
these tasks on the available resources. On the contrary, through compile-time
information, each computational unit of \parsec
immediately releases the dependencies of the completed task solely using the
local knowledge of the DAG. At last, while \parsec uses an opportunistic
approach, the \starpu scheduling strategy exploits cost models of the
computation and data movements to schedule tasks to the right resource
(\cpu{} or \gpu{}) in order to minimize overall execution time.  However, it
does not have a data-reuse policy on \cpu-shared memory systems, resulting in lower
efficiency when no \gpu{}s are used, compared to the data-reuse heuristic of
\parsec.
At the practical level, \parsec supports multiple streams to manage the CUDA
devices, allowing partial overlap between computing tasks, maximizing the
occupancy of the \gpu{}. On the other hand, \starpu allows data transfers
directly between \gpu{}s without going through central memory, potentially
increasing the bandwidth of data transfers when data is needed by multiple
\gpu{}s.

\section{Supernodal factorization over DAG schedulers}
\label{sec:implementation}

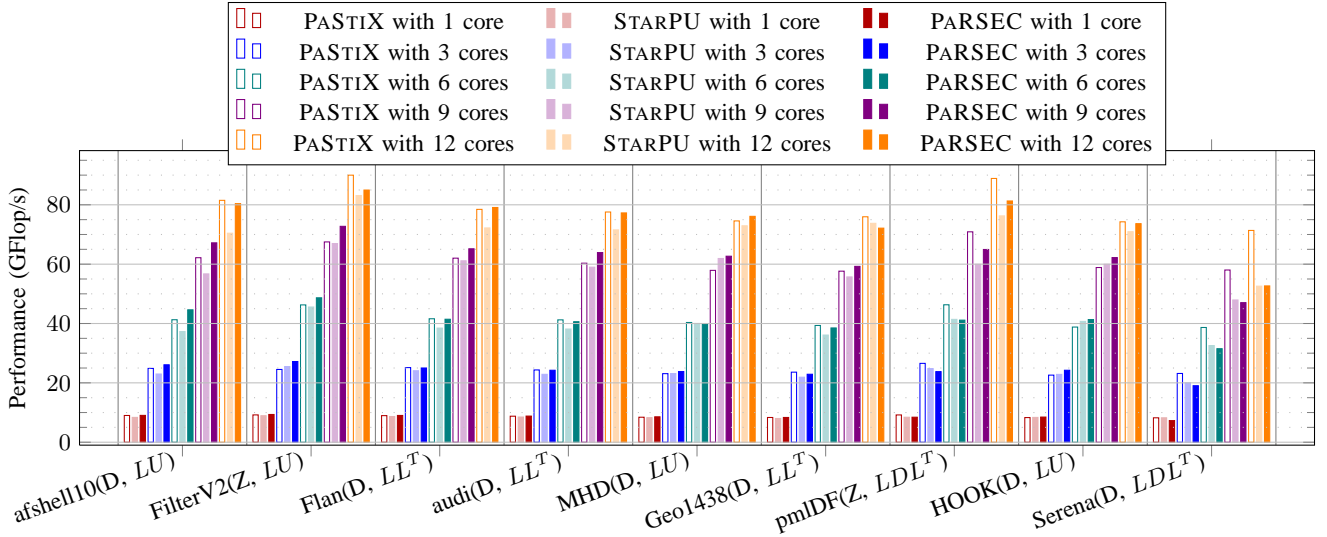
\begin{figure*}[!htbp]
  \begin{center}
\begin{small}
\begin{tikzpicture}
  \pgfplotsset{
    height=5.5cm,
    width=18cm,
    minor y grid style ={loosely dotted},
    major y grid style = thin,
    minor y tick num = 3,
    grid=minor
  }

  \begin{axis}[
      %every axis plot post/.style={/pgf/number format/fixed},
      ymajorgrids,
      yminorticks = true, yminorgrids,
      legend columns=3, 
      barchartstyle,
      bar width=2pt,
    ]

    \addplot+[red, fill=white] table[x=ID, y = GFLOPS1_0] {pastix_results.data};
    \addlegendentry{\pastix{} with 1 core}
    \addplot+[red!30] table[x=ID, y = GFLOPS1_0] {starpu_results.data};
    \addlegendentry{\starpu{} with 1 core}
    \addplot+[red] table[x=ID, y = GFLOPS1_0] {parsec_results.data};
    \addlegendentry{\parsec{} with 1 core}

    \addplot+[blue, fill=white] table[x=ID, y = GFLOPS3_0] {pastix_results.data};
    \addlegendentry{\pastix{} with 3 cores}
    \addplot+[blue!30] table[x=ID, y = GFLOPS3_0] {starpu_results.data};
    \addlegendentry{\starpu{} with 3 cores}
    \addplot+[blue] table[x=ID, y = GFLOPS3_0] {parsec_results.data};
    \addlegendentry{\parsec{} with 3 cores}

    \addplot+[teal, fill=white] table[x=ID, y = GFLOPS6_0] {pastix_results.data};
    \addlegendentry{\pastix{} with 6 cores}
    \addplot+[teal!30] table[x=ID, y = GFLOPS6_0] {starpu_results.data};
    \addlegendentry{\starpu{} with 6 cores}
    \addplot+[teal] table[x=ID, y = GFLOPS6_0] {parsec_results.data};
    \addlegendentry{\parsec{} with 6 cores}

    \addplot+[violet, fill=white] table[x=ID, y = GFLOPS9_0] {pastix_results.data};
    \addlegendentry{\pastix{} with 9 cores}
    \addplot+[violet!30] table[x=ID, y = GFLOPS9_0] {starpu_results.data};
    \addlegendentry{\starpu{} with 9 cores}
    \addplot+[violet] table[x=ID, y = GFLOPS9_0] {parsec_results.data};
    \addlegendentry{\parsec{} with 9 cores}

    \addplot+[orange, fill=white] table[x=ID, y = GFLOPS12_0] {pastix_results.data};
    \addlegendentry{\pastix{} with 12 cores}
    \addplot+[orange!30] table[x=ID, y = GFLOPS12_0] {starpu_results.data};
    \addlegendentry{\starpu{} with 12 cores}
    \addplot+[orange] table[x=ID, y = GFLOPS12_0] {parsec_results.data};
    \addlegendentry{\parsec{} with 12 cores}

    \draw[thin, gray] (axis cs:-0.5,\pgfkeysvalueof{/pgfplots/ymin}) -- (axis cs:-0.5,\pgfkeysvalueof{/pgfplots/ymax});
    \draw[thin, gray] (axis cs:0.5,\pgfkeysvalueof{/pgfplots/ymin}) -- (axis cs:0.5,\pgfkeysvalueof{/pgfplots/ymax});
    \draw[thin, gray] (axis cs:1.5,\pgfkeysvalueof{/pgfplots/ymin}) -- (axis cs:1.5,\pgfkeysvalueof{/pgfplots/ymax});
    \draw[thin, gray] (axis cs:2.5,\pgfkeysvalueof{/pgfplots/ymin}) -- (axis cs:2.5,\pgfkeysvalueof{/pgfplots/ymax});
    \draw[thin, gray] (axis cs:3.5,\pgfkeysvalueof{/pgfplots/ymin}) -- (axis cs:3.5,\pgfkeysvalueof{/pgfplots/ymax});
    \draw[thin, gray] (axis cs:4.5,\pgfkeysvalueof{/pgfplots/ymin}) -- (axis cs:4.5,\pgfkeysvalueof{/pgfplots/ymax});
    \draw[thin, gray] (axis cs:5.5,\pgfkeysvalueof{/pgfplots/ymin}) -- (axis cs:5.5,\pgfkeysvalueof{/pgfplots/ymax});
    \draw[thin, gray] (axis cs:6.5,\pgfkeysvalueof{/pgfplots/ymin}) -- (axis cs:6.5,\pgfkeysvalueof{/pgfplots/ymax});
    \draw[thin, gray] (axis cs:7.5,\pgfkeysvalueof{/pgfplots/ymin}) -- (axis cs:7.5,\pgfkeysvalueof{/pgfplots/ymax});
    \draw[thin, gray] (axis cs:8.5,\pgfkeysvalueof{/pgfplots/ymin}) -- (axis cs:8.5,\pgfkeysvalueof{/pgfplots/ymax});
  \end{axis}
\end{tikzpicture}
\end{small}
  \end{center}
  \caption{\cpu{} scaling study: \gflops{} performance of the
    factorization step on a set of nine matrices with the three
    schedulers.}
  \label{figure.CPU}
\end{figure*}

Similarly to dense linear algebra, sparse direct factorization relies on three
types of operations: the factorization of the diagonal block (\potrf{}), the
solve on off-diagonal blocks belonging to the same panel (\trsm{}), and the
trailing panels updates (\gemm{}).
Whereas the task dependency graph from a dense Cholesky factorization
\cite{icl:509} is extremely regular, the DAG describing the supernodal method
contains rather small tasks with variable granularity and less uniform ranges of
execution space.  This lack of uniformity makes the DAG resulting from a sparse
supernodal factorization complex, accruing the difficulty to efficiently
schedule the resulting tasks on homogeneous and heterogeneous computing
resources.

The current scheduling scheme of \pastix{} exploits a 1D-block distribution,
where a task assembles a set of operations together, including the tasks
factorizing one panel (\potrf{} and \trsm{}) and all updates generated by this
factorization. However, increasing the granularity of a task in such a way
limits the potential parallelism, and has a growing potential of bounding the
efficiency of the algorithm when using many-core architectures.
To improve the efficiency of the sparse factorization on a multi-core
implementation, we introduced a way of controlling the granularity of the BLAS
operations. This
functionality dynamically splits update tasks, 
 so that the
critical path of the algorithm can be reduced.  In this paper, for
both the \parsec
and \starpu runtimes, we split \pastix tasks into two sub-sets of tasks:
\begin{itemize}
\item the diagonal block factorization and off-diagonal blocks updates,
  performed on one panel;
\item the updates from off-diagonal blocks of the panel to one other panel of
  the trailing sub-matrix.
\end{itemize}
Hence, the number of tasks is bound by the number of blocks in the symbolic
structure of the factorized matrix.

Moreover, when taking into account heterogeneous architectures in the
experiments, a finer control of the granularity of the computational tasks is
needed.
% In order to control the granularity of the computational tasks, the criterion
% used to set the block sizes has to be extended to heterogeneous architectures.
Some references for benchmarking dense linear algebra kernels are described
in~\cite{Volkov2008} and show that efficiency could be obtained on \gpu{}
devices only on relatively large blocks -- a limited number of such blocks can
be found on a supernodal factorization only on top of the elimination tree.
Similarly, the amalgamation algorithm~\cite{A:LaBRI::HRR07}, reused from the
implementation of an incomplete factorization, is a crucial step to obtain
larger supernodes and efficiency on \gpu{} devices.  The default parameter for
amalgamation has been slightly increased to allow up to 12\% more fill-in to
build larger blocks while maintaining a decent level of parallelism.

In the remaining of the paper, we present the extensions to the solver to
support heterogeneous many-core architectures. These extensions were validated
through experiments conducted on \textit{Mirage} nodes from the \plafrim{} cluster
at INRIA Bordeaux. A \textit{Mirage} node is equipped with two hexa-core
Westmere Xeon X5650 (2.67 GHz), 32 GB of memory and 3 Tesla M2070 \gpu{}s.
\pastix{} was built without \texttt{MPI} support using \texttt{GCC 4.6.3},
\cuda{} \texttt{4.2}, \texttt{Intel MKL 10.2.7.041}, and \scotch
\texttt{5.1.12b}. %%\starpu{} version was 1.1-r11396 and \pastix{} revision 4231.
Experiments were performed on a set of nine matrices, all part of the University
of Florida sparse matrix collection~\cite{davis}, and are described in
Table~\ref{table.Matrices}. These matrices represent different
research fields and exhibit a wide range of properties (size, arithmetic,
symmetry, definite problem, etc). The last column reports the number
of floating point operations (\flop) required to factorize those
matrices and used to compute the performance results shown in this section.

\begin{table}[!htbp]
    \scalebox{0.9}{
      \begin{minipage}[c]{1.1\linewidth}
  \begin{center}
        \begin{tabular}{|l|l|l|l|l|l|l|}
          \hline
          Matrix    & Prec   & Method  &  Size     & nnz\subscript{A}   & nnz\subscript{L}       & \tflop{}\\
          \hline
          Afshell10 &  D     & \LU{}   &   1.5e+6  & 27e+6  &     610e+6 &   0.12\\
          FilterV2  &  Z     & \LU{}   &   0.6e+6  & 12e+6  &     536e+6 &   3.6\\
          Flan      &  D     & \LLT{}  &   1.6e+6  & 59e+6  &    1712e+6 &   5.3 \\
          Audi      &  D     & \LLT{}  &   0.9e+6  & 39e+6  &    1325e+6 &   6.5\\
          MHD       &  D     & \LU{}   &   0.5e+6  & 24e+6  &    1133e+6 &   6.6 \\
          Geo1438   &  D     & \LLT{}  &   1.4e+6  & 32e+6  &    2768e+6 &   23\\
          Pmldf     &  Z     & \LDLT{} &   1.0e+6  &  8e+6  &    1105e+6 &   28\\
          Hook      &  D     & \LU{}   &   1.5e+6  & 31e+6  &    4168e+6 &   35\\
          Serena    &  D     & \LDLT{} &   1.4e+6  & 32e+6  &    3365e+6 &   47\\
          \hline
        \end{tabular}
        \caption{Matrix description (Z: double complex, D: double).}
        \label{table.Matrices}
  \end{center}
      \end{minipage}
    }
\end{table}

\vspace{-0.5cm}
\subsection{Multi-core Architectures}

As mentioned earlier, the \pastix{} solver has already been optimized for
distributed clusters of NUMA nodes. We use the current state-of-the-art
\pastix{} scheduler as a basis, and compare the results obtained using the
\starpu{} and \parsec{} runtimes from there.
Figure~\ref{figure.CPU} reports the results from a strong scaling
experiment, where the number of computing resources varies from 1 to 12 cores,
and where each group represents a particular matrix. Empty bars correspond to
the \pastix{} original scheduler, shaded bars correspond to \starpu{}, and filled bars
correspond to \parsec{}. The figure is in \flops{}, and a higher value on the Y-axis represents
a more efficient implementation. Overall, this experiment shows that on a shared
memory architecture the performance obtained with any of the above-mentioned
approaches are comparable, the differences remaining minimal on the target
architecture.

We can also see that, in most cases, the \parsec{} implementation is more
efficient than \starpu{}, especially when the number of cores
increases. \starpu{} shows an overhead on multi-core experiments attributed to
its lack of cache reuse policy compared to \parsec{} and the \pastix internal
scheduler.  A careful observation highlights the fact that both runtimes obtain
lower performance compared with \pastix for \LDLT on both PmlDF and Serena
matrices.  Due to its single task per node scheme, \pastix{} stores
the $DL^T$ matrix in a
temporary buffer which allows the update kernels to call a
simple \gemm{} operation. On the contrary, both \starpu and \parsec
implementations are using a less efficient kernel that performs 
the full $LDL^T$ operation at each update. Indeed, due to the extended set of tasks, the life
span of the temporary buffer could cause large memory overhead.
In conclusion, using these generic runtimes shows similar performance and
scalability to the \pastix{} internal solution on the majority of test cases,
while providing a suitable level of performance and a desirable portability,
allowing for a smooth transition toward more complex heterogeneous
architectures.

\subsection{Heterogeneous Architectures Implementation}

While obtaining an efficient implementation was one of the goals of this
experiment, it was not the major one. The ultimate goal was to develop a
portable software environment allowing for an even transition to accelerators, a
software platform where the code is factorized as much as possible, and where
the human cost of adapting the sparse solver to current and future hierarchical
complex heterogeneous architectures remains consistently low.
Building upon the efficient supernodal implementation on top of DAG based
runtimes, we can more easily exploit heterogeneous architectures.  The \gemm{}
updates are the most compute-intensive part of the matrix
factorization, and it is important that these tasks are offloaded to the \gpu{}.
We decide not to offload the tasks that factorize and update the panel to the
\gpu{} due to the limited computational load, in direct relationship with the
small width of the panels.
It is common in dense linear algebra to use the accelerators for the update
part of a factorization while the \cpu{}s factorize the panel; so from this
perspective our approach is conventional.
However, such an approach combined with look-ahead techniques gives really good
performance for a low programming effort on the
accelerators~\cite{yamazaki2012one}. The same solution is applied in this study,
since the panels are split during the analysis step to fit the classic look-ahead
parameters. 

\label{subsec:kernel}

It is a known fact that the update is the most compute intensive task during a
factorization. Therefore, generally speaking, it is paramount to obtain good
efficiency on the update operation in order to ensure a reasonable level of
performance for the entire factorization. Due to the embarrassingly parallel
architecture of the \gpu{}s and to the extra cost of moving the data back and
forth between the main memory and the \gpu{}s, it is of greatest importance to
maintain this property on the \gpu.

\begin{figure}[!htbp]
  \begin{center}
    \includegraphics[width=0.9\linewidth]{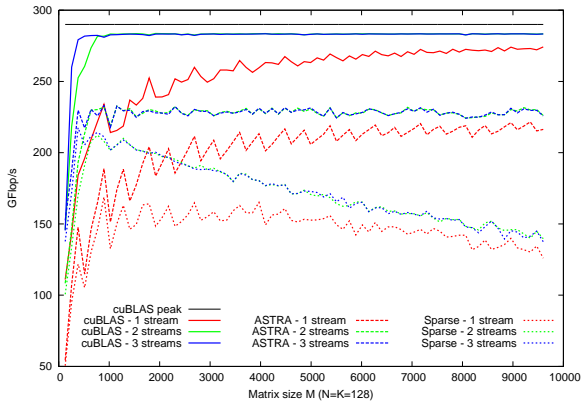}
  \end{center}
  \caption{Multi-stream performance comparison on the \dgemm{}
    kernel for three implementations: \cublas{} library, ASTRA
    framework, and the sparse adaptation of the ASTRA framework.}
  \label{fig:kernel_perf}
\end{figure}

As presented in Figure~\ref{fig:update}, the update task used in the \pastix solver
groups together all outer products that are applied to a same panel. On the \cpu{}
side, this \gemm{} operation is split in two steps due to the gaps in the
destination panel: the outer product is computed in a contiguous temporary
buffer, and upon completion, the result is dispatched on the destination panel.
This solution has been chosen to exploit the performance of vendor provided
\blas{} libraries in exchange for constant memory overhead per working thread.

For the \gpu{} implementation, the requirements for an efficient kernel are
different. First, a \gpu{} has significantly less memory compared with what is
available to a traditional processor, usually in the range of 3 to 6 GB. This
forces us to carefully restrict the amount of extra memory needed during the
update, making the temporary buffer used in the CPU version unsuitable. Second,
the uneven nature of sparse irregular matrices might limit the number of active
computing units per task. As a result, only a partial number of the available
warps on the \gpu{} might be active, leading to a deficient occupancy. Thus, we
need the capability to submit multiple concurrent updates in order to provide
the \gpu{} driver with the opportunity to overlap warps between different tasks
to increase the occupancy, and thus the overall efficiency.

Many CUDA implementations of the dense \gemm{} kernel are available to the
scientific community. The most widespread implementation is provided by Nvidia
itself in the \cublas{} library~\cite{nvidia2008cublas}. This implementation is
extremely efficient since \cuda 4.2 allows for calls on multiple streams, but
is not open source. Volkov developed an implementation for the first generation
of \cuda enabled devices~\cite{Volkov2008} in real single
precision. In~\cite{Tan:2011:FID:2063384.2063431}, authors propose an assembly
code of the \dgemm kernel that provides a 20\% improvement on \cublas{} 3.2
implementation. The \magma library proposed a first implementation
of the \dgemm kernel~\cite{nath2010improved} for the Nvidia Fermi \gpu{}s. Later, an auto-tuned framework, called ASTRA, was
presented in~\cite{10.1109/TPDS.2011.311} and included into the \magma
library. This implementation, similar to the ATLAS library for \cpu{}s, is a
highly configurable skeleton with a set of scripts to tune the parameters for
each precision.

As our update operation is applied on a sparse representation of the panel and
matrices, we cannot exploit an efficient vendor-provided \gemm{} kernel. We
need to develop our own, starting from a dense version and altering the
algorithm to fit our needs.
Due to the source code availability, the coverage of the four floating point
precisions, and it's tuning capabilities, we decided to use the ASTRA-based
version for our \emph{sparse} implementation.
As explained in~\cite{10.1109/TPDS.2011.311} the matrix-matrix operation is
performed in two steps in this kernel. Each block of threads computes the
outer-product $tmp = AB$ into the \gpu shared memory, and then the addition
$C=\beta C + \alpha tmp$ is computed.  To be able to compute directly into $C$,
the result of the update from one panel to another, we extended the kernel to
provide the structure of each panel. This allows the kernel to
compute the correct position directly into $C$ during the \emph{sum} step. This introduces a loss
in the memory coalescence and deteriorates the update parts, however it prevents
the requirement of an extra buffer on the \gpu for each offloaded kernel.

One problem in the best parameters used in the \magma{} library for the ASTRA
kernel is that it has been determined that using textures gives the best
performance for the update kernel. The function \texttt{cudaBindTexture} and
\texttt{cudaUnbindTexture} are not compatible with concurrent kernel calls on
different streams. Therefore, the textures have been disabled in the kernel,
reducing the performance of the kernel by about 5\% on large square matrices.

Figure~\ref{fig:kernel_perf} shows the study we made on the \gemm kernel and the
impact of the modifications we did on the ASTRA kernel. These experiments are done
on a single \gpu of the \emph{Mirage} cluster. The experiments consist of computing
a representative matrix-matrix multiplication of what is typically encountered
during sparse factorization.  Each point is the average performance of 100 calls
to the kernel that computes: $C=C-AB^T$, with $A$, $B$, and $C$,
matrices respectively of dimension $M$-by-$N$, $K$-by-$N$, and $M$-by-$N$. $B$ is taken as the
first block of $K$ rows of $A$ as it is the case in Cholesky factorization.  The
plain lines are the performance of the \cublas library with 1 stream
(\emph{red}), 2 streams (\emph{green}), and 3 streams (\emph{red}). The black
line represents the peak performance obtained by the \cublas library on square
matrices. This peak is never reached with the particular configuration case
studied here.  The dashed lines are the performance of the ASTRA library in the
same configurations. We observe that this implementation already looses
50\gflops, around 15\%, against the \cublas library, and that might be caused by
the parameters chosen by the auto-tuning framework which has been run only on
square matrices.  Finally, the dotted lines illustrate the performance of the
modified ASTRA kernel to include the gaps into the $C$ matrix. For the
experiments, $C$ is a panel twice as tall as $A$ in which blocks are randomly
generated with average size of $200$ rows. Blocks in $A$ are also randomly
generated with the constraint that the rows interval of a block of $A$ is
included in the rows interval of one block of $C$, and no overlap is made
between two blocks of $A$.  We observe a direct relationship between the height
of the panel and the performance of the kernel: the taller the panel, the
lower the performance of the kernel. The memory loaded to do the outer
product is still the same as for the ASTRA curves, but memory loaded for the
$C$ matrix grows twice as fast without increasing the number of \flop to
perform. The ratio \flop per memory access is dropping and explains the
decreasing performance. However, when the factorization progresses and moves up
the elimination trees, nodes get larger and the real number of blocks encountered
is smaller than the one used in this experiment to illustrate worst cases.

Without regard to the kernel choice, it is important to notice how
the multiple streams can have a large impact on the average
performance of the kernel. For this comparison, the 100 calls made in
the experiments are distributed in a round-robin manner over the available
streams. One stream always gives the worst
performance. Adding a second stream increases the performance of
all implementations and especially for small cases
when matrices are too small to feed all resources of the \gpu. The
third one is an improvement for matrices with $M$ smaller than 1000,
and is similar to two streams over 1000.

This kernel is the one we provide to both runtimes to offload
computations on \gpu{}s in case of Cholesky and LU factorizations.
An extension of the kernel has been made to handle the \LDLT{}
factorization that takes an extra parameter to the diagonal matrix $D$
and computes: $C= C - L D L^T$. This modified version decreases the
performance by 5\%.

\subsection{Heterogeneous experiments}
\label{subsec:expes}

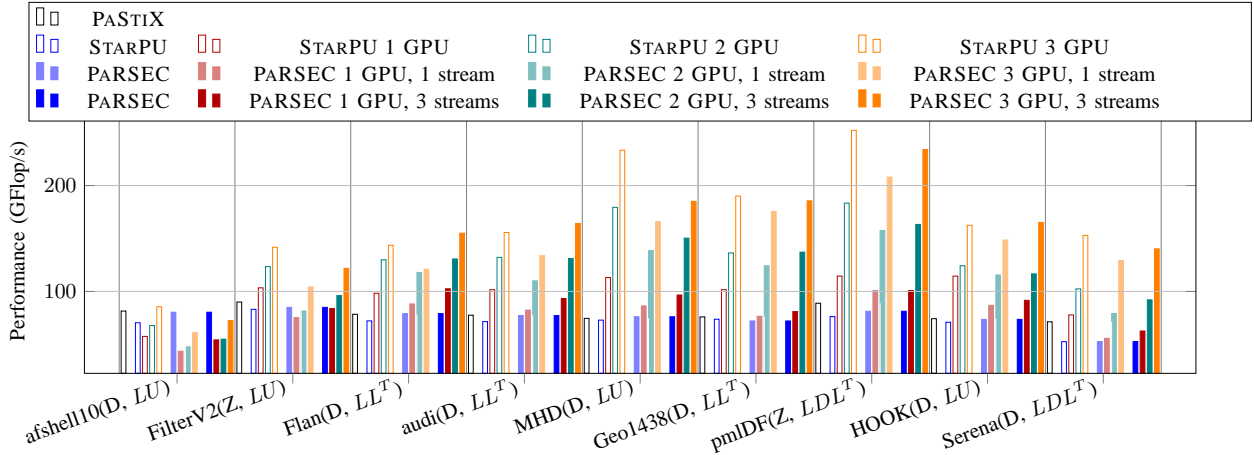
\begin{figure*}[!htbp]
  \begin{center}
    \scalebox{0.9}{
      \begin{minipage}[c]{1.1\linewidth}
\begin{small}
\begin{tikzpicture}
  \pgfplotsset{
    axisstyle/.style={
      compat=1.3,
      legend style={at={(0.7,0.95)}, anchor=north},
      major x tick style = transparent,
      ybar, bar width=4pt,
      ymin=0,
      scaled y ticks = false,
      minor y grid style ={loosely dotted},
      major y grid style = thin,
      minor y tick num = 5,
    }
    grid = minor
  }

  \pgfplotsset{
    height=5.5cm,
    width=18cm
  }

  \begin{axis}[
      legend columns=5,
      ymajorgrids,
      yminorticks = true, yminorgrids,
      barchartstyle
    ]
    \addplot+[black, fill=white] table[x=ID, y = GFLOPS12_0] {pastix_results.data};
    \addlegendentry{\pastix{}}
    \addlegendimage{empty legend}\addlegendentry{}
    \addlegendimage{empty legend}\addlegendentry{}
    \addlegendimage{empty legend}\addlegendentry{}

    \addplot+[white, fill=white, bar width=0pt] table[x=ID, y = GFLOPS12_0] {pastix_results.data};
    \addlegendentry{}
    \addplot+[blue, fill=white] table[x=ID, y = GFLOPS12_0] {starpu_results.data};
    \addlegendentry{\starpu{}}
    \addplot+[red, fill=white] table[x=ID, y = GFLOPS11_1] {starpu_results.data};
    \addlegendentry{\starpu{} 1 \gpu{}}
    \addplot+[teal, fill=white] table[x=ID, y = GFLOPS10_2] {starpu_results.data};
    \addlegendentry{\starpu{} 2 \gpu{}}
    \addplot+[orange, fill=white] table[x=ID, y = GFLOPS9_3] {starpu_results.data};
    \addlegendentry{\starpu{} 3 \gpu{}}

    \addplot+[white, fill=white, bar width=0pt] table[x=ID, y = GFLOPS12_0] {pastix_results.data};
    \addlegendentry{}
    \addplot+[blue!50] table[x=ID, y = GFLOPS12_0] {parsec_results.data};
    \addlegendentry{\parsec{}}
    \addplot+[red!50] table[x=ID, y = GFLOPS12_1] {parsec_results.data};
    \addlegendentry{\parsec{} 1 \gpu{}, 1 stream}
    \addplot+[teal!50] table[x=ID, y = GFLOPS12_2] {parsec_results.data};
    \addlegendentry{\parsec{} 2 \gpu{}, 1 stream}
    \addplot+[orange!50] table[x=ID, y = GFLOPS12_3] {parsec_results.data};
    \addlegendentry{\parsec{} 3 \gpu{}, 1 stream}

    \addplot+[white, fill=white, bar width=0pt] table[x=ID, y = GFLOPS12_0] {pastix_results.data};
    \addlegendentry{}
    \addplot+[blue] table[x=ID, y = GFLOPS12_0] {parsec_results.data};
    \addlegendentry{\parsec{}}
    \addplot+[red] table[x=ID, y = GFLOPS12_1] {parsec_results_3streams.data};
    \addlegendentry{\parsec{} 1 \gpu{}, 3 streams}
    \addplot+[teal] table[x=ID, y = GFLOPS12_2] {parsec_results_3streams.data};
    \addlegendentry{\parsec{} 2 \gpu{}, 3 streams}
    \addplot+[orange] table[x=ID, y = GFLOPS12_3] {parsec_results_3streams.data};
    \addlegendentry{\parsec{} 3 \gpu{}, 3 streams}
    \draw[thin, gray] (axis cs:-0.5,\pgfkeysvalueof{/pgfplots/ymin}) -- (axis cs:-0.5,\pgfkeysvalueof{/pgfplots/ymax});
    \draw[thin, gray] (axis cs:0.5,\pgfkeysvalueof{/pgfplots/ymin}) -- (axis cs:0.5,\pgfkeysvalueof{/pgfplots/ymax});
    \draw[thin, gray] (axis cs:1.5,\pgfkeysvalueof{/pgfplots/ymin}) -- (axis cs:1.5,\pgfkeysvalueof{/pgfplots/ymax});
    \draw[thin, gray] (axis cs:2.5,\pgfkeysvalueof{/pgfplots/ymin}) -- (axis cs:2.5,\pgfkeysvalueof{/pgfplots/ymax});
    \draw[thin, gray] (axis cs:3.5,\pgfkeysvalueof{/pgfplots/ymin}) -- (axis cs:3.5,\pgfkeysvalueof{/pgfplots/ymax});
    \draw[thin, gray] (axis cs:4.5,\pgfkeysvalueof{/pgfplots/ymin}) -- (axis cs:4.5,\pgfkeysvalueof{/pgfplots/ymax});
    \draw[thin, gray] (axis cs:5.5,\pgfkeysvalueof{/pgfplots/ymin}) -- (axis cs:5.5,\pgfkeysvalueof{/pgfplots/ymax});
    \draw[thin, gray] (axis cs:6.5,\pgfkeysvalueof{/pgfplots/ymin}) -- (axis cs:6.5,\pgfkeysvalueof{/pgfplots/ymax});
    \draw[thin, gray] (axis cs:7.5,\pgfkeysvalueof{/pgfplots/ymin}) -- (axis cs:7.5,\pgfkeysvalueof{/pgfplots/ymax});
    \draw[thin, gray] (axis cs:8.5,\pgfkeysvalueof{/pgfplots/ymin}) -- (axis cs:8.5,\pgfkeysvalueof{/pgfplots/ymax});
  \end{axis}
\end{tikzpicture}
\end{small}
      \end{minipage}
    }
  \end{center}
  \caption{\gpu{} scaling study: \gflops{} performance of the
    factorization step with the three schedulers on a set of 10
    matrices. Experiments exploit twelve CPU cores and from zero to
    three additional \gpu{}s.}
  \label{figure.GPU}
\end{figure*}

Figure~\ref{figure.GPU} presents the performance obtained on our set of matrices
on the \textit{Mirage} platform by enabling the \gpu{}s in addition to all
available cores. The \pastix{} run is shown as a reference. \starpu{} runs are empty
bars, \parsec{} runs with 1 stream are shaded and \parsec runs with 3
streams are fully
colored. This experiment shows that we can efficiently use the additional
computational power provided by the \gpu{}s using the generic runtimes.  In its
current implementation, \starpu{} has either \gpu or \cpu worker threads. A \gpu
worker will execute only \gpu tasks. Hence, when a \gpu is used, a \cpu worker
is removed.  With \parsec{}, no thread is dedicated to a \gpu, and they all might
execute \cpu tasks as well as \gpu tasks. The first computational threads that submit
a \gpu{} task takes the management of the \gpu until no \gpu work remains in
the pipeline.
Both runtimes manage to get similar performance and satisfying scalability over
the 3 \gpu{}s. In only two cases, MHD and pmlDF, \starpu{} outperforms \parsec
results with 3 streams.  This experimentation also reveals that, as was
expected, the computation takes advantage of the multiple streams that are
available through \parsec{}. Indeed, the tasks generated by a sparse
factorization are rather small and won't use the entire \gpu{}. This
\parsec feature compensates for the prefetch strategy of \starpu that gave it the
advantage when compared to the one stream results.  One can notice the poor
performance obtained on the \texttt{afshell} test case: in this case, the amount
of \flop{} produced is too small to efficiently benefit from the \gpu{}s.

\section{Conclusion}
\label{sec:conclusion}

In this paper, we have presented a new implementation of a sparse direct solver
with a supernodal method using a task-based programming paradigm.  The programming
paradigm shift insulates the solver from the underlying hardware. The runtime
takes advantage of the parallelism exposed via the graph of tasks to maximize
the efficiency on a particular platform, without the involvement of the
application developer.
In addition to the alteration of the mathematical algorithm to adapt the solver
to the task-based concept, and to provide an efficient memory-constraint sparse
\gemm{} for the \gpu, contributions to both runtimes (\parsec and \starpu) were
made such that they could efficiently support tasks with irregular duration,
and minimize the non-regular data movements to, and from, the devices.
While the current status of this development is already significant in itself,
the existence of the conceptual task-based algorithm opened an astonishing new
perspective for the seamless integration of any type of
accelerator. Providing computational kernels adapted to specialized
architectures has become the only obstruction to a portable, efficient, and
generic sparse direct solver exploiting these devices. In the context of this
study, developing efficient and specialized kernels for \gpu{}s allowed a swift
integration on hybrid platforms.
Globally, our experimental results corroborate the fact that the portability and
efficiency of the proposed approach are indeed available, elevating this
approach to a suitable programming model for applications on hybrid environments.

%However, our study so far barely scratch the improvements that can be added to
%make the integration with the runtime even more smooth and efficient. 
Future work will concentrate on smoothing the runtime integration
within the solver.
First, in order to minimize the scheduler overhead, we plan to increase the
granularity of the tasks at the bottom of the elimination tree. Merging leaves
or subtrees together yields bigger, more computationally intensive tasks.
Second, we will pursue the extension of this work in distributed heterogeneous
environments. On such platforms, when a supernode updates another non-local
supernode, the update blocks are stored in a local extra-memory space (this is
called ``fan-in'' approach~\cite{r97}). By locally accumulating the updates until the last
updates to the supernode are available, we trade bandwidth for latency. The
runtime will allow for studying dynamic algorithms, where the number of local
accumulations has bounds discovered at runtime.
Finally, the availability of extra computational resources highlights the
potential to dynamically build or rebuild the supernodal structures according to
the load on the cores and the \gpu{}s. 

\section*{Acknowledgment}
The work presented in this paper is supported by the Inria Associated
Team MORSE, 
%%to redesign dense
%%and sparse linear algebra methods for maximum efficiency on large scale
%%heterogeneous multi-core architectures.  This material is based upon work
%%supported
by the National Science Foundation under Grant Number CCF-1244905,
and by the ANR ANEMOS ANR-11-MONU-002. The authors would like to
thanks E.~Agullo, G.~Boutin, and A.~Guermouche.

%% The authors would like to thank the \dague\ and \starpu\ teams for
%% their support and assistance with this project. Special thanks also go to
%% Sam~Crawford for the valuable comments on our 
%% submitted manuscript and to Abdou~Guermouche for his advice.

%%%%%%%
%
%  Bibliography
%
%%%%%%%




%%\newcommand{\BIBdecl}{\setlength{\itemsep}{0.2\baselineskip}} 
\bibliographystyle{IEEEtran}
\bibliography{icl,pastix,all}



% \end{thebibliography}

%%\IEEEtriggeratref{8}
\end{document}